\documentclass[
aps,%
12pt,%
final,%
notitlepage,%
oneside,%
onecolumn,%
nobibnotes,%
nofootinbib,
superscriptaddress,%
noshowpacs,%
centertags]%
{revtex4}

\begin{document}
\selectlanguage{english} 
\title{NeXSPheRIO Results on Elliptic-Flow Fluctuations at RHIC}
\author{\firstname{Yogiro} \surname{Hama}}
\affiliation{Instituto de F\y sica, Universidade 
             de S\~ao Paulo, C.P. 66318, 05315-970 
             S\~ao Paulo-SP, Brazil}
\author{\firstname{Rone Peterson} \surname{G. Andrade}}
\affiliation{Instituto de F\y sica, Universidade 
             de S\~ao Paulo, C.P. 66318, 05315-970 
             S\~ao Paulo-SP, Brazil}

\author{\firstname{Fr\'ed\'erique} \surname{Grassi}}
\affiliation{Instituto de F\y sica, Universidade 
             de S\~ao Paulo, C.P. 66318, 05315-970 
             S\~ao Paulo-SP, Brazil}

\author{\firstname{Wei-Liang} \surname{Qian}}
\affiliation{Instituto de F\y sica, Universidade 
             de S\~ao Paulo, C.P. 66318, 05315-970 
             S\~ao Paulo-SP, Brazil}

\author{\firstname{Takeshi} \surname{Osada}}
\affiliation{Musashi Institute of Technology, 
             Tokyo, Japan}

\author{\firstname{Carlos Eduardo} \surname{Aguiar}}
\affiliation{Instituto de F\y sica, Universidade 
             Federal do Rio de Janeiro, C.P. 68528, 
             21945-970 Rio de Janeiro-RJ , Brazil}

\author{\firstname{Takeshi} \surname{Kodama}}
\affiliation{Instituto de F\y sica, Universidade 
             Federal do Rio de Janeiro, C.P. 68528, 
             21945-970 Rio de Janeiro-RJ , Brazil}

\begin{abstract} 
By using the NexSPheRIO code, we study the elliptic-flow  
fluctuations in Au+Au collisions at $200\,A\,$GeV. 
It is shown that, by fixing the parameters of the model 
to correctly reproduce the charged pseudo-rapidity and 
the transverse-momentum distributions, reasonable 
agreement of $<v_2>$ with data is obtained, both as 
function of pseudo-rapidity as well as of transverse 
momentum, for charged particles. Our results on elliptic-flow fluctuations are in good agreement with 
the recently measured data on experiments. 
\end{abstract} 
\maketitle
\section{Introduction}
It is by now widely accepted that hydrodynamics is a 
successful approach for describing the collective 
flow in high-energy nuclear collisions. 
The basic assumption in hydrodynamical models is the 
local thermal equilibrium. It is assumed that, after a complex process involving microscopic collisions of nuclear constituents, at a certain early instant a hot 
and dense matter is formed, which would be in local 
thermal equilibrium. After this instant, the system 
would evolve hydrodynamically, following the well 
known set of differential equations. 

However, since our systems are not large enough, important event-by-event fluctuations are expected. With regard to this question, fluctuation in the initial conditions deserves a special consideration. Because the incident nuclei are not smooth objects, if thermalization is verified at very early 
time as usually assumed in hydrodynamic approach, the initial conditions for hydrodynamics, expressed by distributions of velocity and thermodynamic quantities at such an instant, could not be smooth and would fluctuate from event to event. 

In the past few years, we have studied several effects  caused by such fluctuating and non-smooth initial conditions on some observables, by using a code especially developed for this purpose, which we call NeXSPheRIO \cite{1,2,3,4,5}.
~In particular, we showed in preliminary works on Au+Au 
collisions at $130A\,$GeV that fluctuations of $v_2$ are 
quite large~\cite{1}.~
Recently,\hfilneg\ 

\ni $\sigma_{v_2}/\langle v_2\rangle$ data were obtained in  $200\,A\,$GeV Au+Au collisions \cite{6,7,8},
~showing a good agreement with our previous results, when  
QGP is included. The main object of this communication is 
to check it at the correct energy and with a more updated 
version of the code.  

In what follows, we will first give a brief description of the NeXSPheRIO code. This will be done in the next  Section. Then, we explain in Section~\ref{parameters} 
how this code is used to compute the observables of our 
interest. We show the results of computations in 
Section~\ref{results} where effects of fluctuations in 
the initial conditions are emphasized. Finally, 
conclusions are drawn and further outlook are given. 

\section{NeXSPheRIO Code}
\label{NS} 
NeXSPheRIO is a junction of two codes: NeXus and  SPheRIO. 
The NeXus code \cite{9}
~is used to compute the initial conditions (IC) $T^{\mu \nu}$, $j^{\mu}$ and  $u^{\mu}$ on some initial hypersurface. It is a 
microscopic model based on the Regge-Gribov theory and 
the main advantage for our purpose is that, once 
a pair of incident nuclei or hadrons and their incident 
energy are chosen, it can produce, in the  event-by-event basis, detailed space distributions of  energy-momentum tensor, baryon-number, strangeness and charge densities, at a given initial time $\tau=\sqrt{t^2-z^2}\sim1\,$fm. 
Remark that, when we use a microscopic model to create 
a set of IC for hydrodynamics, the energy-momentum  tensor produced by the microscopic model does not 
necessarily correspond to that of local equilibrium, so 
we need to transform it to that of the equilibrated  
matter, adopting some procedure as described in detail 
in Ref. \cite{10}.~

\begin{figure*}[t!] 
\setcaptionmargin{5mm} 
\onelinecaptionsfalse 
\includegraphics[angle=-90, width=15.cm]{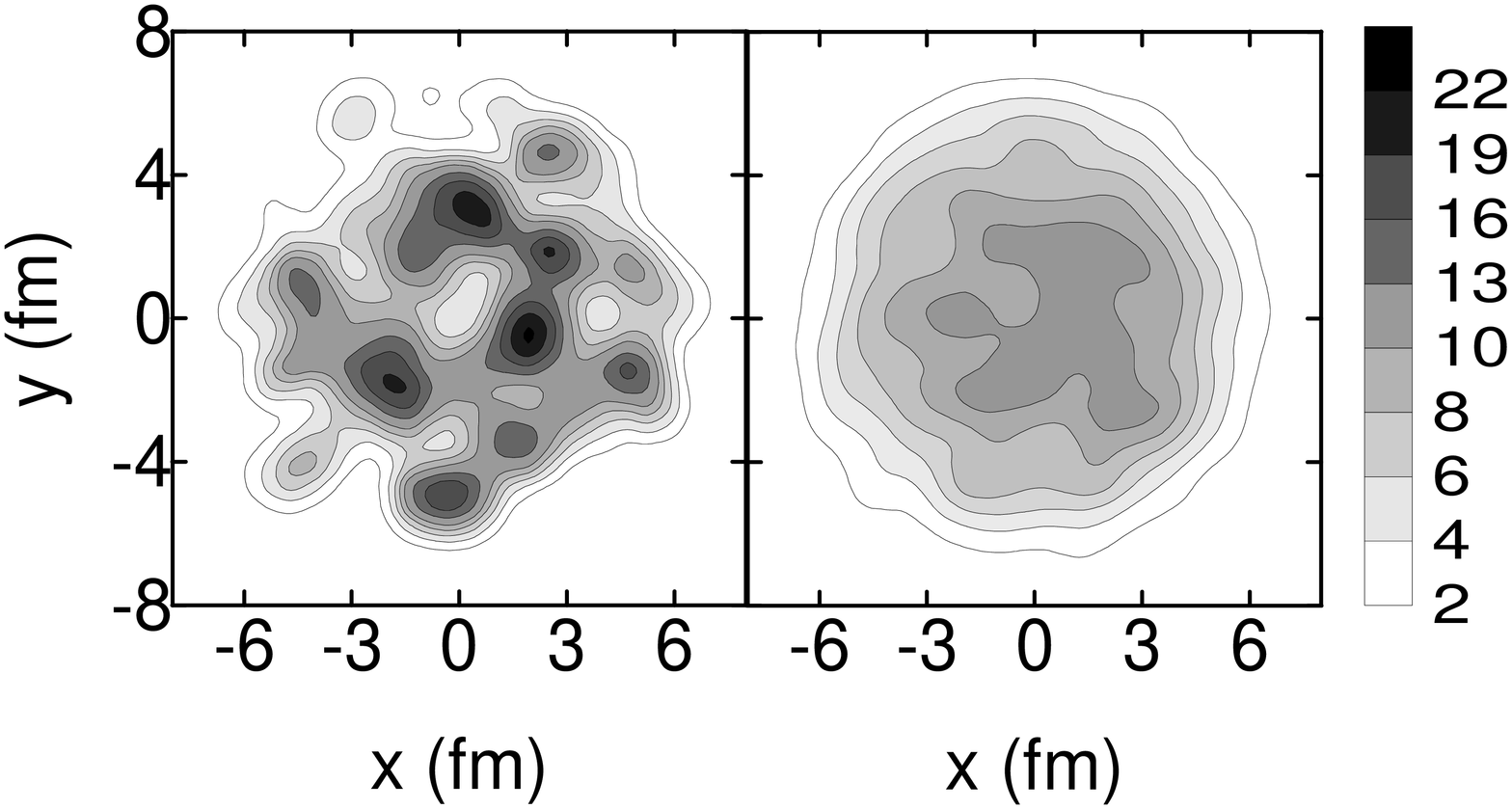} 
\captionstyle{normal} 
\caption{Examples of initial conditions for central Au+Au 
collisions given by NeXus at mid-rapidity plane. The 
energy density is plotted in units of GeV/fm$^3$. 
Left: one random event. Right: average over 30 random 
events (corresponding to the smooth initial conditions 
in the usual hydro approach).}
\label{ic} 
\end{figure*}

We show in Fig.$\,$\ref{ic} an example of such a 
fluctuating event, produced by NeXus event generator, 
for central Au~+~Au collision at $130\,A\,$GeV,  compared with an average over 30 events. As can be seen, the 
energy-density distribution for a single event (left), 
at the mid-rapidity plane, presents several blobs of 
high-density matter, whereas in the averaged IC (right) the distribution is smoothed out, even though the number of events is only 30. The latter would corresponds to the usually adopted smooth and symmetrical IC in many hydrodynamic calculations. 
The bumpy event structure, as exihibited in  
Fig.$\,$\ref{ic}, was also shown in calculations with 
{HIJING}$\,$\cite{11}.
~As already observed there and studied 
in$\,$\cite{1,2,3,4,5,10} 
this bumpy structure gives important consequences in 
the observables. 

Solving the hydrodynamic equations for events, so 
irregular as the one shown in Fig~\ref{ic}, 
requires a special care. The SPheRIO code is well 
suited to computing the hydrodynamical evolution of 
such systems. It is based on Smoothed Particle 
Hydrodynamics (SPH), a method originally developed in 
astrophysics \cite{12}
~and adapted to relativistic heavy ion collisions \cite{13}.
~It parametrizes the flow in terms of discrete Lagrangian 
coordinates attached to small volumes (called 
``particles'') with some conserved quantities. Its main 
advantage is that any geometry in the initial conditions 
can be incorporated and giving a desired precision. 

Now, we have to specify some equation of state (EoS) 
describing the locally equilibrated matter. Here, 
in accordance with Ref.~\cite{4},
~we will adopt a phenomenological implementation of EoS, 
giving a critical end point in the QGP-hadron gas 
transition line, as suggested by the lattice 
QCD~\cite{14}.

Although too simplified, we shall neglect in the following any dissipative effects and also assume the 
usual sudden freezeout at a constant temperature. 
As for the conserved quantities, besides the energy,  momentum and entropy, we consider just the baryon  number. 

In computing several observables, NeXSPheRIO described  here is run many times, corresponding to many different 
events or initial conditions. At the end, an average 
over final results is performed. We believe that this 
mimics more closely the experimental conditions, as 
compared to the canonical approach where usually smooth 
initial conditions for just one (averaged) event are 
adjusted to reproduce some selected data.

\section{Adjusting the Parameters of the Model} 
 \label{parameters} 
Having been depicted our tool, let us now explain 
how we fix the parameters of the model and compute the  
observables of our interest. 

First of all, since it is impossible to know the impact 
parameter in experiments, we use some quantity which in 
principle can be experimentally determined to define 
the centrality. For instance, in our code, it is possible to determine the number of participant nucleons in each  
event which is intimately connected to the often used ZDC 
energy. Although the participant number is closely 
related to the impact parameter, it is not the 
same \cite{3}~
due to fluctuations. 

\begin{figure*}[t!] 
\setcaptionmargin{5mm} 
\onelinecaptionsfalse 
\includegraphics[width=15.cm]{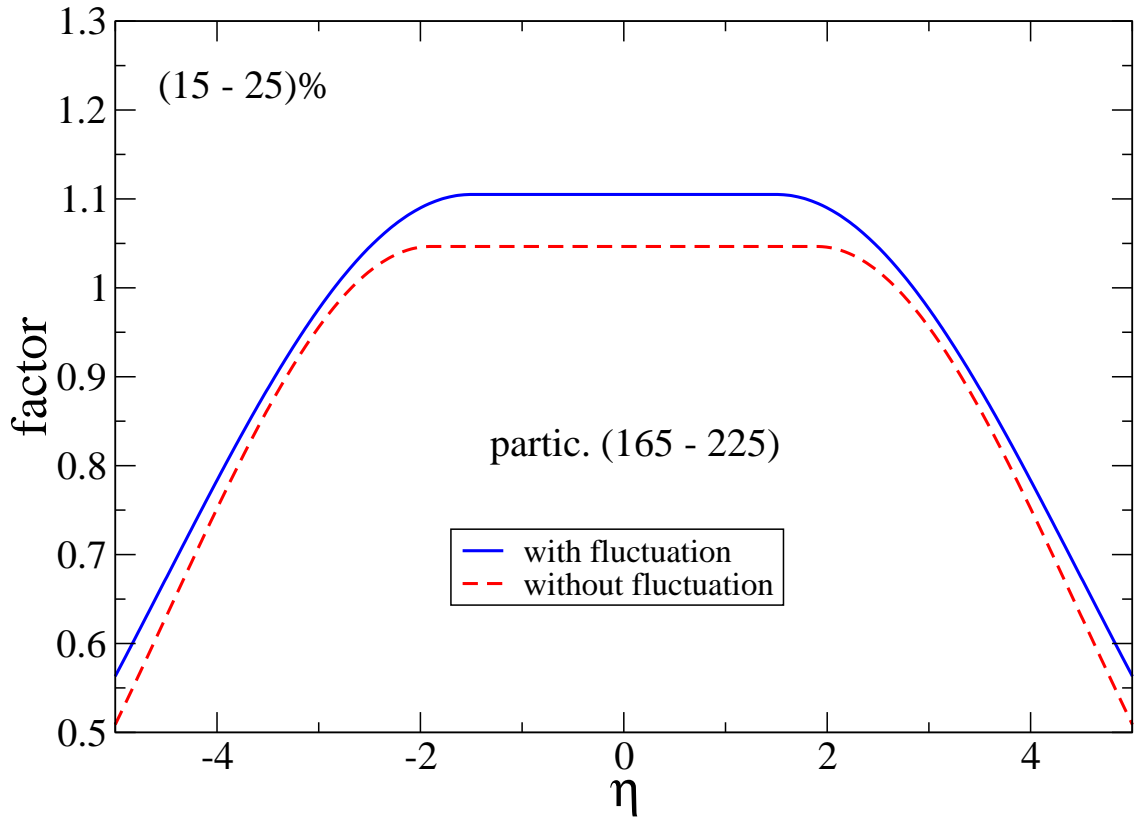}
\captionstyle{normal}
\caption{Example of $\eta$-dependent factor which was 
used to multiply the energy density given by NeXus in the initial conditions of each event.} 
\label{rf}
\end{figure*}

\begin{figure*}[t!] 
\setcaptionmargin{5mm} 
\onelinecaptionsfalse 
\includegraphics[width=15.cm]{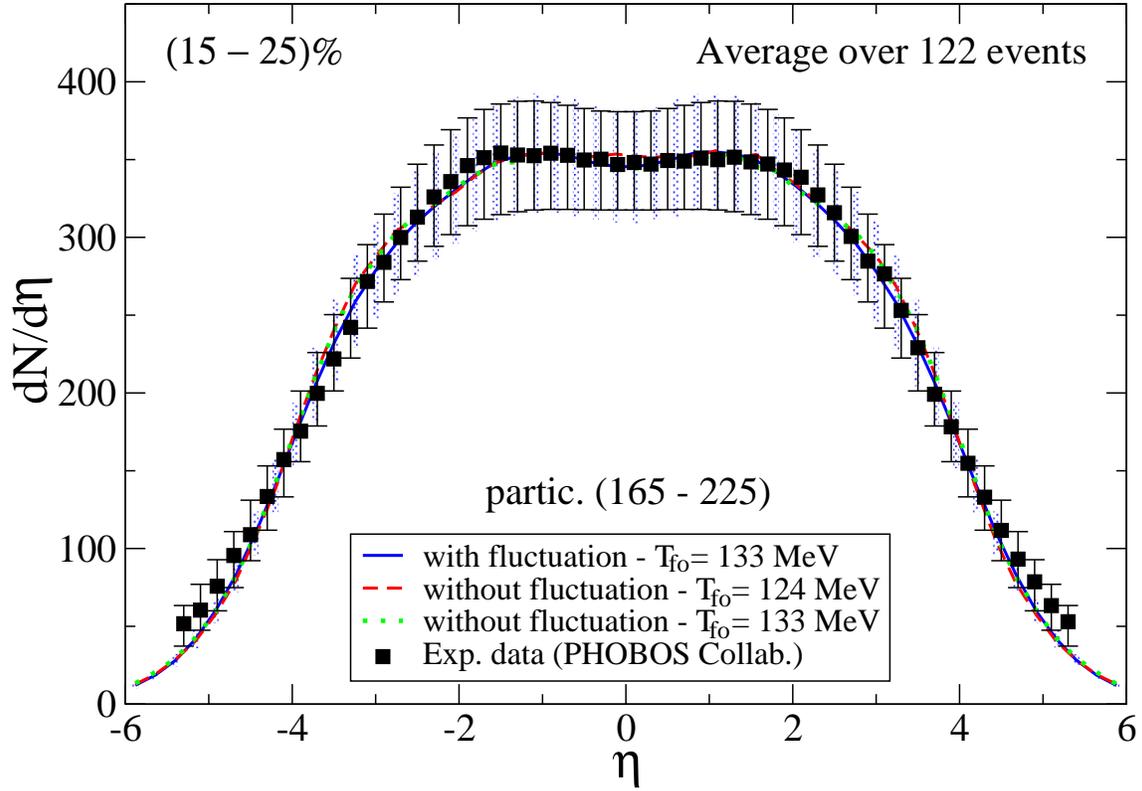}
\captionstyle{normal}
\caption{Results of pseudo-rapidity distributions 
calculated with the NeXus initial conditions as explained in the text. PHOBOS data \cite{15}~
are shown for comparison.} 
\label{dndeta3}
\end{figure*}

Now, certainly any model to be considered as such should reproduce the most fundamental, global  quantities involving the class of phenomena for which 
it is proposed. So, we begin by fixing the initial 
conditions so as to reproduce properly the 
(pseudo-)rapidity distributions of charged particles 
in each centrality window. 
This is done by applying an $\eta$-dependent factor 
$\sim1$ to the initial energy density distribution of all the events of each centrality class, produced by NeXus. Examples of such factors are shown in 
Fig.~\ref{rf} for the centrality (15 - 25)\% class  
both for fluctuating IC and the averaged IC.  
We show the resultant pseudo-rapidity distributions in 
Fig.~\ref{dndeta3}. Here, {\it without fluctuation} 
means that the computation has been done for one event 
whose IC are the average of the same 122 fluctuating 
IC, used in the other case, except for the  
normalization factor shown in Fig.~\ref{rf}. Results 
with averaged IC are being shown in comparison, to 
clearly exhibit the effects of fluctuating initial 
conditions. 
Observe that, to obtain the same multiplicity as shown 
in Fig.~\ref{dndeta3}, we have to start with a smaller 
average energy density in the case of averaged IC, as 
implied by the normalization factor of Fig.~\ref{rf}. 
This is a manifestation of the effect already discussed 
in Ref.~\cite{10}.~
Another observation concerning Fig.~\ref{dndeta3} is 
that the freezeout temperature, $T_{fo}\,$, gives a negligible influence on the (pseudo-)rapidity distributions. 

\begin{figure*}[t!] 
\setcaptionmargin{5mm} 
\onelinecaptionsfalse 
\includegraphics[width=15.cm]{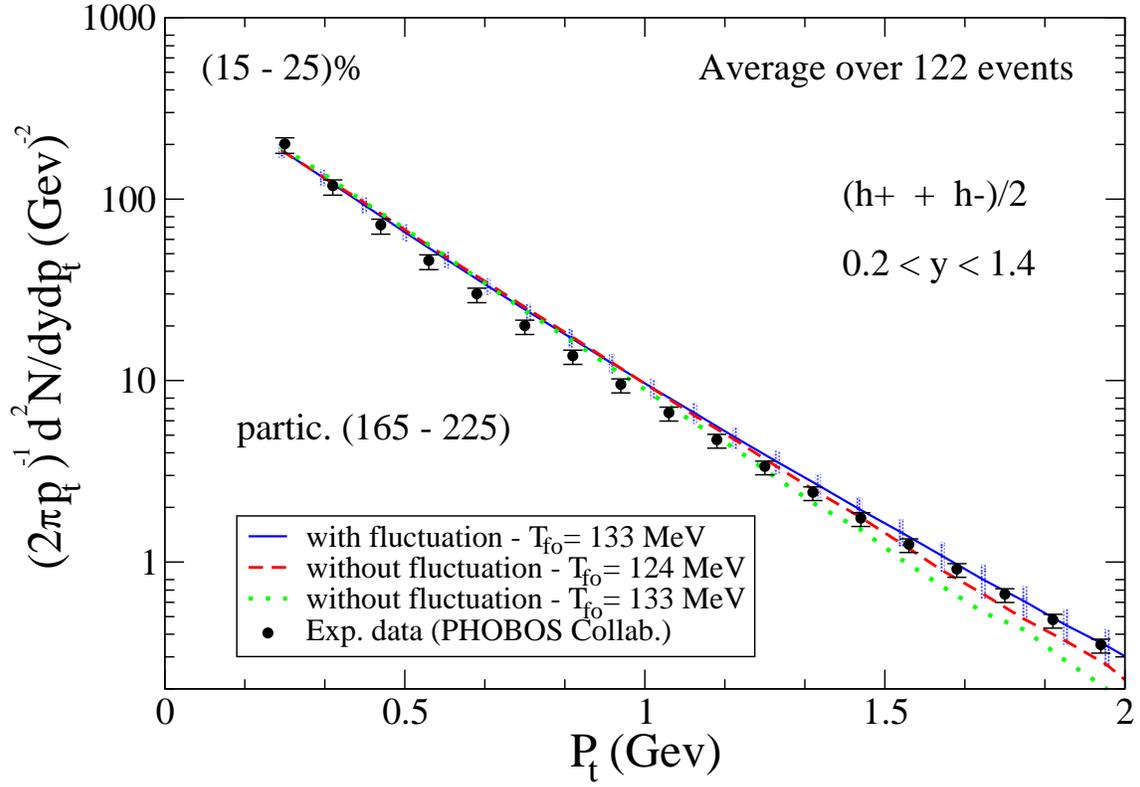}
\captionstyle{normal}
\caption{Results of transverse-momentum distributions 
calculated with the NeXus initial conditions as  explained in the text. PHOBOS data \cite{16}~
are shown for comparison.} 
\label{dndpt3}
\end{figure*}

Next, we would like to correctly reproduce the  transverse-momentum spectra of charged particles, 
which can be achieved by choosing an appropriate 
freezeout temperature, $T_{fo}\,$. Figure~\ref{dndpt3} 
shows examples of choice with the corresponding 
spectra. One can see in this Figure that the 
fluctuating IC make the transverse-momentum spectra 
more concave, closer to data. Also, one sees that 
higher freezeout temperature is required in this case, 
as compared to the one for averaged IC. These 
characteristics are consequences of the bumpy structure 
of the energy-density distribution, as shown in 
Fig.~\ref{ic}, because, those high-energy spots produce 
higher acceleration than the smooth distribution, due 
to a higher pressure gradient. 

\section{Results} 
 \label{results} 

\begin{figure*}[t!] 
\setcaptionmargin{5mm} 
\onelinecaptionsfalse 
\includegraphics[width=15.cm]{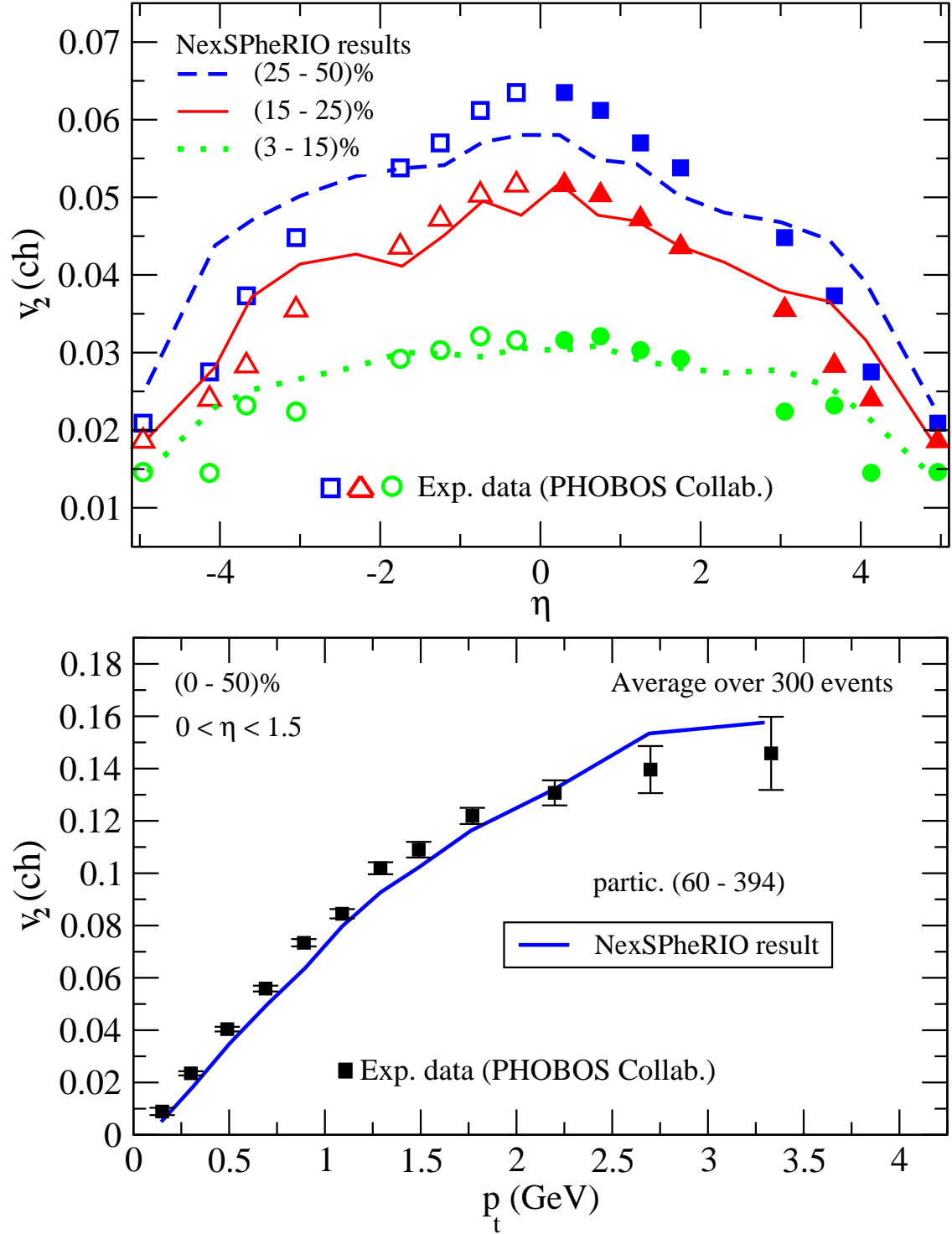}
\captionstyle{normal}
\caption{$v_2$ computed as explained in the text, compared with 
PHOBOS data \cite{17}.~
The upper panel shows the $\eta$ distributions in three 
centrality windows as indicated. The lower panel shows the 
$p_T$ distribution at the mid-rapidity region. 
} 
\label{v2}
\end{figure*}

In Fig.~\ref{v2}, upper panel, we show the pseudo-rapidity 
distributions of $v_2$ for charged particles, calculated  
in three centrality windows as indicated. It is seen that they reasonably reproduce the overall  behavior of the existing data, both the centrality 
and the $\eta$ dependences. The lower panel of Figure~\ref{v2} 
shows the transverse-momentum distribution of $v_2$ in the mid-rapidity region. Again, the main feature is very well reproduced. Notice that, differently from the usual hydrodynamic calculations, the curve shows some bending 
at large-$p_T$ values that is due, in our opinion, to 
the granular structure of our initial conditions, which 
produces a violent isotropic expansion at the beginning, 
so reducing the anisotropy of large-$p_T$ components. 
This question is being studied more carefully. 

\begin{figure*}[t!] 
\setcaptionmargin{5mm} 
\onelinecaptionsfalse 
\includegraphics[width=15.cm]{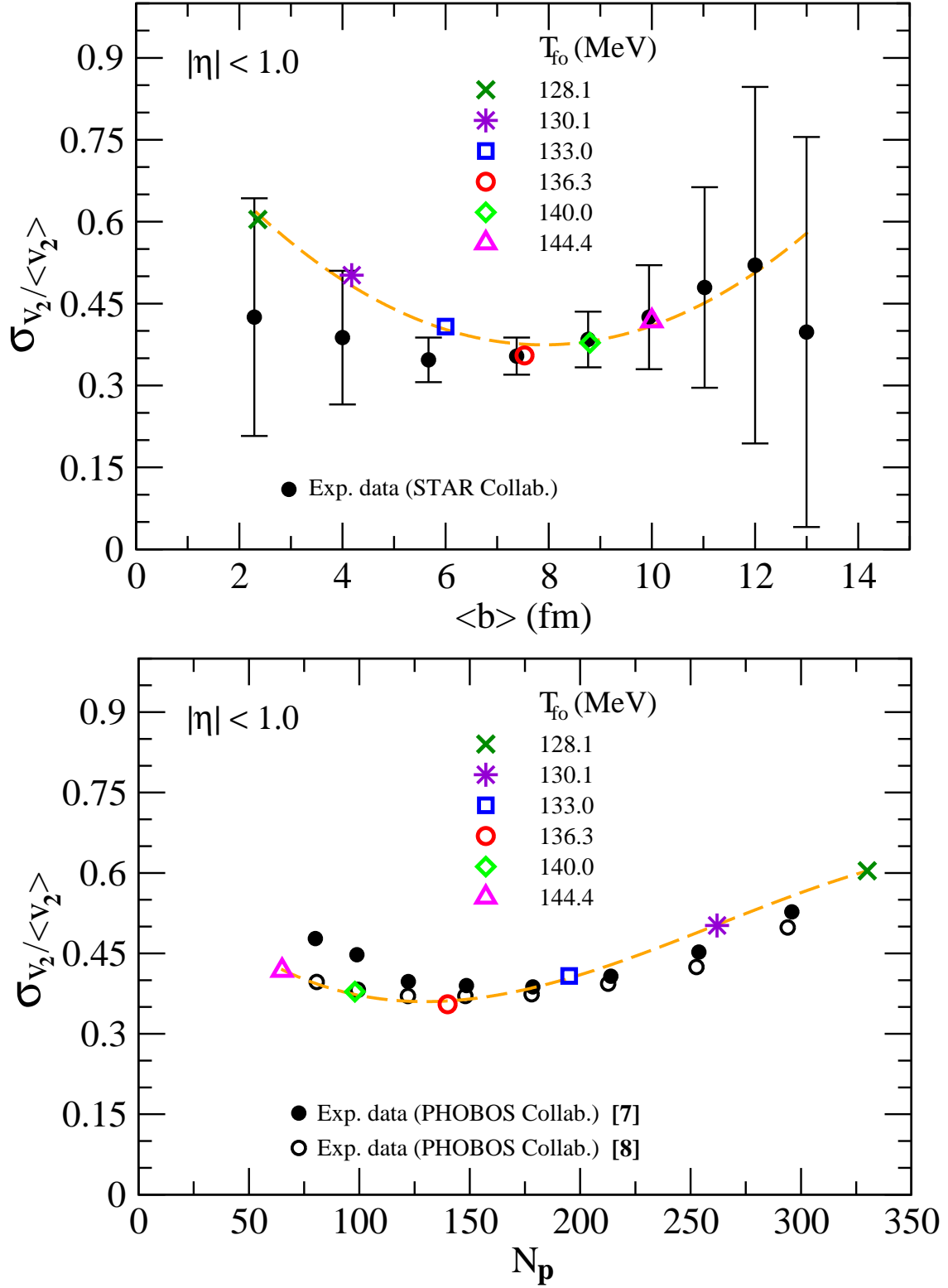}
\captionstyle{normal}
\caption{$\sigma_{v_2}/\langle v_2\rangle$ computed for 
Au+Au collisions at $200\,A\,$GeV, compared with data. 
In the upper panel, $\sigma_{v_2}/\langle v_2\rangle$ is given 
as function of  
the impact parameter $\langle b\rangle$ and compared with the 
STAR data \cite{6}.~
In the lower panel, the same results are expressed as function 
of participant nucleon number $N_p$ and compared with the 
PHOBOS data, \cite{7,8}.
} 
\label{flv2}
\end{figure*}

Now, we show the results for $v_2$ fluctuations in 
Fig.~\ref{flv2}. The freeze-out temperature has been  
chosen as explained in the previous Section and increases 
with the impact parameter $\langle b\rangle$ (decreases 
with the participant nucleon number $N_p$), or the fluid 
decouples hotter and hotter as one goes from more central 
to more peripheral collisions as expected. We remark 
that, also in computing the $p_T$ distribution of $v_2$ 
shown in Fig.~\ref{v2}, these values of temperature have 
been used and then averaged over the partial windows. 
The curves of $v_2$ fluctuations, plotted in 
Fig.~\ref{flv2}, indicate that the NeXSPheRIO results  remain in nice  agreement with the data, also in the  
present calculation.

\section{Conclusions} 
 \label{conclusions} 

In this communication, we gave an account of a check 
we made of the previous results on charged $v_2$  fluctuations~\cite{1},~
to see whether a more careful computation at the correct 
energy $200\,A\,$GeV and with a more updated version of 
the code can still reproduce the recently measured  $\sigma_{v_2}/\langle v_2\rangle$ data~\cite{6,7,8}.

In our model, the fluctuations of the observables appear mostly because of the initial condition fluctuations, 
introduced by the NeXus generator~\cite{9},~
with some additional small effects appearing from the 
freeze-out procedure with Monte-Carlo method. The latter 
is, however, usually made negligible with increasing 
Monte-Carlo events at the freezeout. 

Our conclusion is that we are in the correct way, being 
able to reproduce the essential features of $v_2$ for 
charged particles, and our previous prediction for $v_2$ 
fluctuations, with QGP introduced, remain valid for 
$200\,A\,$GeV Au+Au collisions. 

As mentioned in the previous Section, we are further 
studying in more detail effects of inhomogeneity of the initial conditions on $v_2\,$. Another study in progress is the effects of continuous emission instead of 
sudden freezeout.

\newpage
%

\newpage
%

\newpage
%

\newpage
%

\newpage
%

\newpage
%


\newpage
\begin{thebibliography}{99}

\bibitem{1} 
\refitem{preliminary} T. Osada, C.E. Aguiar, Y. Hama, 
 and T. Kodama, in Proc. of the 6th. RANP Workshop, 
 eds. T. Kodama {\it et al.}, World Scientific, 
 Singapore (2001), pg. 174 [nucl-th/0102011]; 
\refitem{pre2} C.E. Aguiar, Y. Hama, T. Kodama and 
 T. Osada, Nucl. Phys. A{\bf 698}, 639c (2002). 

\bibitem{2} 
\refitem{hbt} O. Socolowski Jr., F. Grassi, Y.~Hama 
 and T. Kodama, Phys. Rev. Lett., {\bf 93}, 182301 (2004); 
\refitem{hbt2} Y.~Hama, F. Grassi, O. Socolowski Jr. and 
 T. Kodama, Acta Phys. Pol. {\bf B36}, 347 (2005). 

\bibitem{3} 
\refitem{b_part} C.E. Aguiar, R. Andrade, F. Grassi, 
 Y. Hama, T. Kodama, T. Osada and O. Socolowski Jr., 
 Braz. J. Phys. {\bf 34}, 319 (2004). 

\bibitem{4} 
\refitem{qm05} Y. Hama, R.P.G. Andrade, F. Grassi, 
 O. Socolowski Jr., T. Kodama, B. Tavares and 
 S.S. Padula, QM05 proceedings, Nucl. Phys. A{\bf 774}, 
 169 (2006). 

\bibitem{5} 
\refitem{v2} R. Andrade, F. Grassi, Y. Hama, T. Kodama, 
 O. Socolowski Jr., Phys. Rev. Lett. {\bf 97}, 202302 
 (2006); 
\refitem{v2_2} ISMD07 Proceedings, Braz. J. Phys. 
 {\bf 37}, 717 (2007). 
 
\bibitem{6} 
\refitem{STAR} P. Sorensen [STAR Collab.], presented at 
 QM 2006, nucl-ex/0612021. 

\bibitem{7} 
\refitem{PHOBOS} C. Loizides [PHOBOS Collab.], 
 presented at QM 2006, nucl-ex/0701049.  

\bibitem{8} 
\refitem{PHOBOS2} B. Alver {\em et al.}, 
 nucl-ex/0702036. 

\bibitem{9} 
\refitem{nexus} H.J. Drescher, F.M. Liu, 
 S. Ostrapchenko, T. Pierog and K. Werner, Phys. Rev.  
 C{\bf 65}, 054902 (2002). 

\bibitem{10} 
\refitem{review} Y. Hama, T. Kodama and 
 O. Socolowski Jr., Braz. J. Phys. {\bf 35}, 24 (2005). 
 
\bibitem{11} 
\refitem{gyulassy} M. Gyulassy, D.H. Rischke and 
 B. Zhang, Nucl. Phys. {\bf A613}, 397 (1997). 
 
\bibitem{12} 
\refitem{sph} L.B. Lucy, Ap. J. {\bf 82}, 1013 (1977); 
\refitem{sph2} R.A. Gingold and J.J. Monaghan, 
 Mon. Not. R. Astr. Soc. {\bf 181}, 375 (1977). 
  
\bibitem{13} 
\refitem{sp1} C.E. Aguiar, T. Kodama, T. Osada, 
 Y. Hama, J. Phys. G{\bf 27}, 75 (2001). 
 
\bibitem{14} 
\refitem{LQCD} Z. Fodor and S.D. Katz, J. High Energy 
 Phys. {\bf 03} (2002) 014; 
\refitem{LQCD2} F. Karsh, Nucl. Phys. A{\bf 698}, 199 
 (2002); 
\refitem{LQCD3} S.D. Katz, QM05 proceedings, Nucl. Phys. 
 A{\bf 774}, 159 (2006). 

\bibitem{15} 
\refitem{PHOBOS3} PHOBOS Collab., B.B. Back 
 {\em et al.}, Braz. J. Phys. {\bf 34}, 829 (2004). 
 
\bibitem{16} 
\refitem{PHOBOS4} PHOBOS Collab., B.B. Back {\em et al.}, 
 Phys. Lett. B{\bf 578}, 297 (2004). 

\bibitem{17} 
\refitem{PHOBOS5} PHOBOS Collab., B.B. Back {\em et al.}, 
 Phys. Rev. C{\bf 72}, 051901 (2005).   
 
\end{thebibliography}
\end{document}